\documentclass[structabstract]{aa}
\usepackage{graphicx}
\usepackage{txfonts}
\usepackage{amssymb}

\begin{document}
   
   \title{Hard MeV-GeV spectra of blazars}
   
   
   \author{K. Katarzy\'nski}

   \offprints{Krzysztof Katarzy\'nski \\kat@astro.uni.torun.pl}

   \institute{
              Toru\'n Centre for Astronomy, Nicolaus Copernicus University, 
              ul. Gagarina 11, PL-87100 Toru\'n, Poland
             }

   \date{Received 3 March 2011 / Accepted 11 October 2011}

   \abstract{}            
            {
             Very high energy (VHE) gamma-ray emission from a distant source 
             ($z \gtrsim 0.2$) can be efficiently absorbed my means of the electron-positron 
             pair creation process. Analyses of the unabsorbed spectra imply that 
             the intrinsic TeV emission of some blazars is hard, with spectral indices
             $0.5 < \alpha < 1$. The absorption depends on the level of 
             extragalactic background light (EBL) that is difficult to measure
             directly. This implies that it is difficult to estimate the slope 
             of the intrinsic TeV emission. To test our blazar emission scenario 
             that is capable to reproducing the hard spectra, we therefore used the 
             observations made by the Fermi Gamma-ray Space Telescope in the 
             unabsorbed MeV-GeV energy range. 
            }
            {
             We assume that the X-ray and gamma-ray emission of TeV blazars is produced in
             a compact region of a jet uniformly filled by particles of relatively 
             high energy ($\gamma\gtrsim 10^3, E=\gamma m_{\rm e} c^2)$. In other words,
             we assume a low energy cut-off in the particle energy distribution.
             The emission produced by the particles with this energy spectrum can 
             explain hard intrinsic spectra in the energy range from MeV up to TeV. 
             We demonstrate how to estimate the basic physical parameters of a source in 
             this case and how to explain the observed spectra by a precise simulation 
             of the particle energy evolution. 
            } 
            {
             To test our estimation methods, we use the observations of two blazars with 
             exceptionally hard spectral indices ($\alpha \lesssim 0.5$) in the MeV-GeV range
             and known redshifts: RGB J0710+591 and 1ES 0502+675. 
             The estimated values of the Doppler factor and magnetic field are 
             compared with our numerical simulations, which confirm that the particle energy 
             distribution with a low energy cut--off can explain the observed hard spectra well. 
             In addition, we demonstrate that the radiative cooling caused by the inverse-Compton 
             emission in the Klein-Nishina regime may help us to explain the hard spectra.
            }
            {}
            {}

\titlerunning{Hard MeV-GeV spectra of blazars}
\authorrunning{Katarzy\'nski}   
\maketitle

\section{Introduction}

The emission of some blazars is observed from radio frequencies up 
to VHE gamma rays. Collating observations from different energy ranges, 
we can show that the spectra of these objects contain two characteristic
peaks (in $\nu F(\nu)$ plots). The first peak appears in the X-rays from
a few keVs to a few hundred keVs and the second peak is observed 
around an energy of a few TeVs (e.g. Ghisellini, G. et al. 
\cite{Ghisellini98}, Massaro et al. \cite{Massaro04}, Nieppola et al. 
\cite{Nieppola06}, Ghisellini, G. et al. \cite{Ghisellini10}, 
Abdo et al. \cite{Abdo10b}). Since the TeV emission is the most 
prominent feature in those sources we often call them TeV 
blazars, which constitutes a relatively small group of sources 
within the blazar family. The gamma-ray emission of most blazars 
peaks in the MeV-GeV range. Moreover, TeV gamma rays 
are efficiently absorbed by the EBL. This in addition limits 
the number of observed TeV blazars.

The high energy emission of blazars is believed to
originate inside their jets. However, the observed variability on
time scales from days (e.g. Catanese et al. \cite{Catanese97},
Fossati et al. \cite{Fossati08}) down to a few minutes (e.g.
Aharonian et al. \cite{Aharonian07a}) indicates that only
a small part of a jet is radiating at high energies.
This kind of emission requires high energy particles,  
which can gain energy by acceleration at the
front of a shock wave inside the jet. This is the simplest explanation
of the acceleration process, which assumes that a small fraction
of the jet bulk kinetic energy is transferred to the particles.
In the presence of a magnetic field, the particles can 
radiate this energy through synchrotron emission, generating
the first peak in the spectrum. Some fraction of
the synchrotron photons can then be up-scattered to higher energies.
This is the inverse-Compton (IC) scattering that gives the second
peak in the TeV energies. This simple scenario has been proposed many times 
as part of models of the VHE emission of blazars (e.g. Dermer et al. 
\cite{Dermer97}, Bloom \& Marscher \cite{Bloom96}, Inoue \& Takahara 
\cite{Inoue96}, Mastichiadis \& Kirk \cite{Mastichiadis97},
Katarzy\'nski et al. \cite{Katarzynski01}).

A TeV gamma ray photon travelling through the intergalactic medium can
interact with an infra-red photon producing an electron-positron pair. 
This causes absorption of the emission above a few hundred 
GeVs. To calculate this absorption, we have to determine the 
level of the intergalactic radiation field, which is difficult 
to measure directly but can be estimated from simulations 
of star light production in evolving galaxies. There are many 
different solutions for the level of the absorption
(e.g. De Jager \& Stecker \cite{DeJager02}, Keneiste et al. 
\cite{Kneiske04}, Franceschini et al. \cite{Franceschini08}, 
Kneiske, \& Dole \cite{Kneiske10}). Thus the slope of the 
intrinsic TeV emission cannot be accurately calculated. Nevertheless,
the TeV emission of some relatively distant sources
(e.g. 1ES 1101-232 $z= 0.186$, Aharonian et al. \cite{Aharonian06} or 
1ES 0229+200, $z=0.14$, Aharonian et al. \cite{Aharonian07b})
indicates that the intrinsic spectra are hard ($F \propto 
\nu^{-\alpha}$ with $0.5 < \alpha < 1$), even if we assume as low
as possible absorption. 

It is difficult to explain how these hard spectra could be 
created. Most of the standard emission models assume a power-law
or broken power-law particle energy distribution with a 
slope  $n \simeq 2$ (where the number of particles $N \propto \gamma^{-n}$) 
and a minimum energy equivalent to $\gamma_{\rm m} \simeq 1$. This gives 
the spectral index of the synchrotron emission $\alpha=(n-1)/2=0.5$. 
This emission is up-scattered to higher energies by the
same population of the particles, hence the index of
the inverse-Compton spectrum should be similar to the
index of the synchrotron emission. Note that high energy
particles ($\gamma \gtrsim 10^5$) do not scatter efficiently 
because of the Klein-Nishina (KN) restrictions. The broken 
power-law particle distribution ($n_1$ and $n_2$ below and above 
the break, respectively) implies that there is a broken synchrotron 
spectrum ($\alpha_1$ and $\alpha_2$), for which the spectral index in the KN regime 
is approximately $\alpha \simeq 2 \alpha_2 - \alpha_1 = 2.5$ 
for $n_2=4$ (Tavecchio et al \cite{Tavecchio98}).

A question to ask is how we can explain the hard spectra from MeVs up 
to TeV gamma rays. One of the simplest solution is to assume a
cut-off in the low energy part of the electron spectrum, which
basically means $\gamma_{\rm m} >> 1$. This solution was proposed
for the first time by Katarzynski et al. (\cite{Katarzynski06a})
to explain the observations of 1ES 1101-232 in the TeV range. 
This cut-off leads to an additional break in the synchrotron spectrum
between IR and X-ray energies. The spectral index of the first 
part of the spectrum, from radio up to X-ray energies, 
becomes a constant $\alpha_0 = -1/3$. This is a ``tail" of 
the synchrotron emission produced by the low energy 
particles. This part of the radiation field with such a
hard spectral index can also be up-scattered. Therefore,
the IC spectrum can also be very hard up to limiting value 
$\alpha = -1/3$. Simultaneous optical and X-ray observations 
of 1ES 0229+200 performed by the {\it Swift} satellite (Tavecchio et al. 
\cite{Tavecchio09}) shows an abrupt break between the optical 
and the X-ray range. This strongly implies that there is a low energy 
cut-off to the particle energy distribution.

In the present work, we explore this idea of a low energy cut-off.
We focus on the unabsorbed MeV-GeV range where the spectral
index of the intrinsic emission is observed directly. We 
demonstrate how to estimate the basic parameters of a source and 
how to explain the observed spectra by a simulation 
of the particle energy evolution. We apply our estimations 
and modelling to the X-ray and gamma-ray observations of two
blazars with known redshifts. 

\section{Observations}

The spectra of TeV blazars in the MeV-GeV range are always hard 
with spectral indices $\alpha < 1$. The average expected value of 
the spectral index in this particular range is $\alpha \sim 0.5$. 
In this work we focus on the sources with $\alpha < 0.5$ that 
we call exceptionally hard.

For all blazars observed by the Fermi Gamma-ray Space Telescope
(Atwood et al. \cite{Atwood09}) from 20 MeV up to 300 GeV, only 
in some cases were spectra published in the Fermi Large Area 
Telescope First Source Catalog (Abdo et al. \cite{Abdo10a}) 
exceptionally hard with the photon index $\Gamma \lesssim 1.5$ 
(where the photon index is greater by unity in comparison to 
the spectral index). Moreover, in many cases we either do not know 
the distance to the observed sources (e.g. RBS 158 $\Gamma=1.34 
\pm 0.33$, RBS 621 $\Gamma=1.42 \pm 0.32$) or the spectra were 
obtained with a relatively large uncertainty (e.g. 1ES 1101-232 
$\Gamma=1.36 \pm 0.58$). Therefore, we selected only two 
objects for our test. 

The observational parameters of the first object RGB J0710+591 
(z=0.125) in the {\it Fermi} catalogue are  $\Gamma=1.28 
\pm 0.22$, and $F = (1.46 \pm 0.49)$ $10^{-11}$ erg cm$^{-2}$ s$^{-1}$, 
where $F$ is the energy flux in the 100 MeV to 100 GeV range. This 
source was also discovered by the VERITAS gamma-ray 
observatory in the TeV range (Acciari et al. \cite{Acciari10}). 
The photon index obtained by VERITAS above $\sim 300$ GeV was
$\Gamma=2.69 \pm 46$. In our analysis. we also use X-ray observations 
obtained by the {\it Swift} satellite in the energy range from 0.2 
to 10 keV (Acciari et al. \cite{Acciari10}). A notable characteristic
of this object is that the X-ray peak is observed at the level 
a few times higher than the GeV-TeV peak. This indicates that 
the synchrotron cooling is dominating in this source.

The photon index of the second source 1ES 0502+675 (z=0.341) in 
the Fermi catalogue is rather soft $\Gamma=1.75\pm0.11$. However,
the MeV-GeV spectrum of this source is quite complex. Owing
the limited photon statistics, the observations published 
by Abdo et al. (\cite{Abdo10b}) and  Abdo et al. (\cite{Abdo10c}) 
seem to display a broken power-law 
spectrum of IC emission. This is rather unexpected 
and not predicted by any model of blazar emission. However, 
this may have a simple explanation if we assume for example 
that two independent jet components were observed at the 
same time. The break in the spectrum appears around an energy of 1 GeV. 
Fitting a single power--law function to the observations
above this energy, we obtained $\Gamma=1.35\pm0.20$ and 
$\Gamma=1.46\pm0.18$ for the data sets above quoted, respectively.
The VERITAS team reported the detection of this source in the TeV energy
range (Ong \cite{Ong09}). However, no public information about 
the observed flux and slope has been publicly available so far.
In addition we use quasi-simultaneous observations from 
{\it Swift} (Abdo et al. \cite{Abdo10b}). The comparison between 
the X-ray peak level and the GeV observations shows that in this 
particular source the radiative cooling generated by the IC 
emission is the predominant cooling process.

\section{Estimations}

To estimate the basic parameters of a source of VHE emission, we adapted
approach proposed by Tavecchio et al. (\cite{Tavecchio98}). We
assume a spherical source of radius $R$, filled uniformly by 
a constant magnetic field ($B$) and relativistic particles ($K$ -- density). 
In addition, we assume that the particle energy distribution is a broken 
power-law ($n_1, n_2, \gamma_{\rm b}$ -- break energy) and that the source 
travels with relativistic velocity ($\delta$ -- Doppler factor)
inside a jet. Finally, we included a low energy cut-off 
in the particle energy spectrum ($\gamma_{\rm m}$ -- minimum energy). 
This is our main extension to the initial approach that let us
estimate the basic physical parameters from the exceptionally hard spectrum. 
Even this simple scenario has eight free parameters. Therefore, 
using observational quantities we can only estimate the relation
between the main parameters $B$ and $\delta$. However, we can
estimate this relation in four different ways.

\begin{figure}
\centering
\includegraphics[width=9cm]{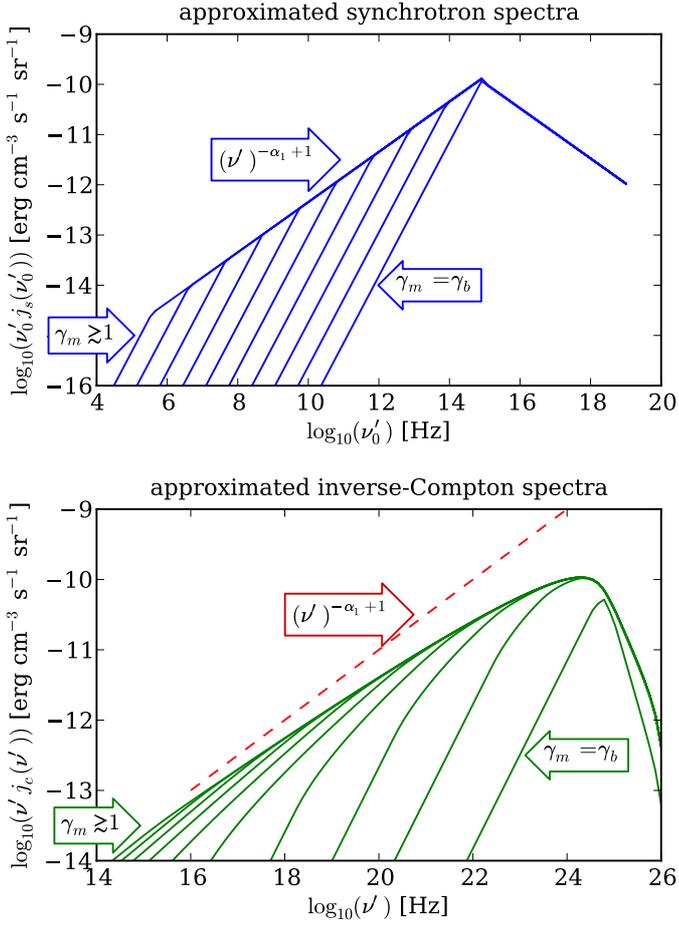}
\caption{Approximate synchrotron and IC spectra calculated for
different values of $\gamma_m$, from almost unity up to $\gamma_b$. 
This test demonstrates how the spectral index in the MeV-GeV range
depends on the $\gamma_m$ value. The dashed line shows the difference
between the estimated ($\alpha_1=0.5$) and calculated ($ \alpha \simeq 0.65$) 
spectral slope below the peak.
}
\label{fig_spe_ser}
\end{figure}

\subsection{Peak positions}

The simplest estimation of the $B - \delta$ relation involves
the comparison of the synchrotron peak position ($\nu_{\rm s}$) with
the inverse-Compton peak frequency ($\nu_{\rm c}$). The positions 
of the peaks do not depend on $\gamma_{\rm m}$ thus we can directly 
use the formula derived by Tavecchio et al. (\cite{Tavecchio98})
\begin{equation}
B = (1+z)\frac{\nu_{\rm s}^2}{2.8 \times 10^6 \nu_{\rm c}} \delta^{-1}.
\label{eq_est_pek}
\end{equation}
In Fig. \ref{fig_spe_ser}, we illustrate how an approximate IC 
spectrum evolve with time for different values of $\gamma_m$, 
from $\gamma_m \gtrsim 1$ up to $\gamma_b$. The spectra are 
calculated according to the prescription given in the appendix.
This simple test shows that the IC peak remains at the same 
position for the wide range of $\gamma_m$ values with 
the exception for $\gamma_m \simeq \gamma_b$ and this gives 
the limit to the above formula. The IC spectrum with 
$\alpha = -1/3$ that could be obtained for $\gamma_m = \gamma_b$ 
was never observed directly. The observed spectral indices 
described in the previous section are in range from 0.3 to 0.5. 
Therefore, we can safely use this $B$ and $\delta$ relationship 
for our estimations.

The spectra in Fig. \ref{fig_spe_ser} were calculated
for $n_1=2$ and $n_2=4$, which directly gives the spectral
indices of the synchrotron emission below ($\alpha_1 = 0.5$)
and above ($\alpha_2=3/2$) the peak, respectively. The
value of $n_1=2$ is rather typical for many different
astrophysical sources. This may be the result of the 
first--order Fermi acceleration at a shock front or 
the synchrotron or the inverse-Compton cooling 
in the Thompson regime. This index is crucial for 
the emission of TeV blazars because this part of the 
particle spectrum provides synchrotron photons for 
the IC scattering. The IC spectrum in the MeV-GeV 
range comes mostly from the scattering in the Thompson
regime. Therefore, the spectral index of this 
emission should be similar to the index of the 
scattered radiation field. However, the complexity
of the scattering (see Fig. \ref{fig_com_emi}) 
produces a curved spectrum below the peak. In the
MeV-GeV range, just below the peak, the spectrum
can be approximated well by a power-law function
with the index $\alpha \simeq 0.65$. This differs 
significantly from the estimated value $\alpha = 0.5$
that is clearly illustrated in Fig. \ref{fig_spe_ser}.
Note that this concerns only the case where $\gamma_m 
\sim 1$ because $\gamma_m$ controls the slope 
in the other cases. Since the somewhat classical 
value of $n_1=2$, postulated by many different 
particle evolution scenarios, gives $\alpha=0.65$
all spectra with $\alpha<0.65$ should be 
classified as exceptionally hard. Using this revised 
criterion, all spectra considered here may be regarded
as exceptionally hard.

\subsection{Peak levels}

The second relationship between $B$ and $\delta$ derived
by Tavecchio et al. (\cite{Tavecchio98}) was obtained 
from the well--known formula
\begin{equation}
\frac{U_{\rm syn}}{U_B} = \frac{L_c}{L_s},
\end{equation}
where $U_{\rm syn}$ and $U_B$ are the energy densities of
the synchrotron radiation field and the magnetic field,
respectively, and $L_s$ and $L_c$ are the total luminosities
of the synchrotron and the IC emission, respectively. The 
luminosities were calculated from the observed spectral
slopes and the peak emission levels. Unfortunately,
we cannot adopt this approach in our estimations because
usually we do not observe the break in the synchrotron
emission ($\nu_m$) that is related to $\gamma_m$. We instead
compare
\begin{equation}
\frac{j_c(\nu'_c)}{j_s(\nu'_s)} = \frac{F_c(\nu_c)}{F_s(\nu_s)}
\end{equation}
the emissivities (Eq. \ref{eq_syn_emi}, Eq. \ref{eq_com_emi}) with
the observed emission levels at the peaks. This gives the particle 
density (Eq. \ref{eq_bpl_esp}) for the first part of the particle 
energy distribution
\begin{equation}
K_1 = \frac{(x'_s)^{-\alpha_1} F_c(\nu_c)}{j_{c, {\rm sim}}(x'_c) F_s(\nu_s)},
\end{equation}
where
\begin{eqnarray}
j_{c, {\rm sim}}(x') & = & \frac{\sigma_T R}{4} \left\{ \left(\frac{3 x'}{4} \right)^{-\alpha_1} \ln\left(\frac{\gamma_b^2}{(x')^2}\right) \right.\nonumber\\
& + & \left. \frac{\gamma_b^{n_2-n_1}}{\alpha_2-\alpha_1} \left(\frac{3 x'}{4} \right)^{-\alpha_2} 
\left(\frac{3 x'}{4 \gamma_b}\right)^{\alpha_2-\alpha_1} \right\}
\end{eqnarray}
is the simplified version of the IC emissivity calculated as a sum of two
dominant components $j_{c, {\rm sim}} = j_{(1,1)} + j_{(1,2)}$ (Eqs. \ref{eq_com_aa}, 
\ref{eq_com_ab}). Note that in the component $j_{(1,2)}$ the lower integration 
boundary was neglected.
Finally, using the synchrotron emissivity (Eq. \ref{eq_syn_emi}) and the
transformation given by Eq. \ref{eq_flx_tra} we can derive another relation
between $B$ and $\delta$
\begin{equation}
B = \left\{ \frac{\frac{1}{3} C(n_1) (1+z) R^3 K_1 \nu_s^{-\alpha_1}}{F_s(\nu_s) D_L^2} \right\}^\frac{1}{-1-\alpha_1}
\delta^\frac{3}{-1-\alpha_1},
\label{eq_est_lev}
\end{equation}
where the luminosity distance ($D_L$) is calculated in a $\Lambda$CDM 
universe for $h=0.72$, $\Omega_{\Lambda} = 0.7$, and $\Omega_m=0.3$ for
all estimations and the spectra calculated in this work.

\subsection{Spectral index in MeV-GeV range}

Another constraint on the basic physical parameters can be obtained 
from the slope of the gamma-ray emission in the MeV-GeV range. This 
constraint assumes that the particle minimal energy ($\gamma_m$) is 
directly related to the radiative cooling inside the source. This is
one of the main assumptions in our model that we discuss in detail in
the next section. According to this assumption
\begin{equation}
\gamma_m = \frac{3 m_e c^2}{4 \sigma_T c t_{\rm cool} (U_B + U_{\rm syn})},
\label{eq_gam_min}
\end{equation}
where $t_{\rm cool}$ is the characteristic cooling time equivalent
to the source evolution time in this particular case. This and
the synchrotron energy density ($U_{\rm syn}$) are unknown parameters. 
Therefore, for our estimation we assume that 
$U_{\rm syn} = x_{B} U_B$ is energy independent and can be 
parametrized as some fraction ($x_B$) of $U_B$. In the same 
way, we parametrize $t_{\rm cool} = x_R R/c$ as the multiplied 
crossing time. Since the spectral peaks are 
observed at almost the same level we may conclude that
$x_B \sim 1$. In addition, we know that $x_R$ must be greater
than unity, although, in our estimations we assume that $x_R \ge 2$. 
This gives the shortest reasonable time for the 
source evolution. With the above assumptions, we can derive
an upper limit to the magnetic field value
\begin{equation}
B \le \sqrt{\frac{3 m_e c^2 \pi}{2 \sigma_T R \gamma_m}},
\end{equation}
where $\gamma_m$ value required by the above formula can be obtained
from the spectral slope in the MeV-GeV range. However, there is
no simple analytic formula that gives $\gamma_m$.

We derive the the minimum energy ($\gamma_m$) generating 
spectra from $\gamma_m = 1$ up to $\gamma_m = \gamma_b$ and
searching for the required spectral index
\begin{equation}
\alpha = \frac{\ln\left(j_c(\nu'_2)/ j_c(\nu'_1)\right)}{\ln\left(\nu'_1/\nu'_2\right)},
\end{equation}
where $\nu'_1$ and $\nu'_2$ are the corresponding frequencies to the energies
100 MeV and 100 GeV, respectively, in the comoving frame. The frequency 
transformation, which is required for this calculation, introduces
the Doppler factor dependency into the estimation.
Note that it is necessary to use the full emissivity (Eq. \ref{eq_com_emi}) 
for this particular estimation. The emissivity 
can be calculated for any non-zero value of $R, B,$ and $K$ 
because the absolute level of the emission is unimportant 
in this calculation. Finally, to calculate the emissivity
we need the energy spectrum break that is given by a simple 
formula
\begin{equation}
\gamma_b = x'_c \exp\left( \frac{1}{2(\alpha_2 - \alpha_1)} - \frac{1}{\alpha_1 -1} \right)
\end{equation}
derived by Tavecchio et al. (\cite{Tavecchio98}).

\subsection{Pair absorption}

The optical depth for the pair absorption inside a spherical source
can be approximated as
\begin{equation}
\tau(x'_0) = \frac{1}{5} \sigma_T n_s(x'_0) x'_0 2 R,
\end{equation}
where $n_s = 3 \pi j_s R / (h c x'_0)$ is the number density of
soft photons per energy interval (e.g. Coppi \& Blandford \cite{Coppi90}).
The above formula incorporates a simple relation between the soft photon 
and the gamma-ray photon energy $x'_0 = 1/x'$. Using the flux 
transformation (Eq. \ref{eq_flx_tra}), we can write
\begin{equation}
\tau = \frac{9}{10} \frac{\sigma_T F_s(\nu_0) D_L^2}{\delta^3 (1+z) R h c},
\end{equation}
which should be smaller than unity for an optically thin source. This gives a
lower limit to the value of the Doppler factor
\begin{equation}
\delta \ge \left[ \frac{9}{10} \frac{\sigma_T (1+z)^{2 \alpha_1-1} F_s(\nu_0) D^2_L}{ R h c }  \right]^{\frac{1}{2 \alpha_1 + 3}},
\label{eq_par_lim}
\end{equation}
where $\nu_0 = (m_e c^2/h)^2/\nu_c$. Note that this formula is more accurate
for a spherical source than the relationship derived by Dondi \& Ghisellini 
(\cite{Dondi95}) used in the paper by Tavecchio et al. (\cite{Tavecchio98}).
Our formula infers a lower limit that is almost five times higher than the 
above quoted relationship.

\begin{figure}
\centering
\includegraphics[width=9cm]{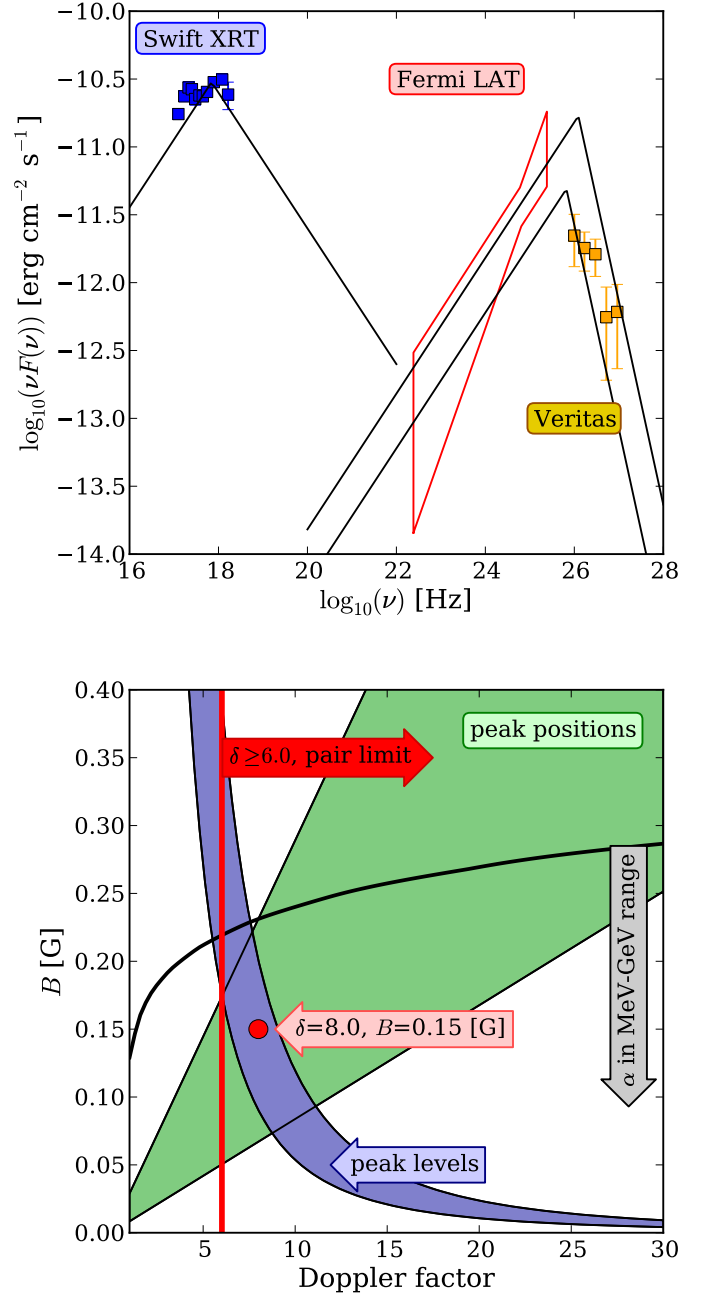}
\caption{The upper panel shows the high energy emission of RGB J0710+591 and approximated spectra 
         used for our estimations. In the lower panel, we illustrate the constraints to the basic 
         parameters obtained in four different ways and the values selected for the modelling.
        }
\label{fig_rgb_est}
\end{figure}

\subsection{Estimation results}

It is difficult to determine the position of the peaks in RGB J0710+591
from the available observations. The {\it Swift} energy range is quite 
narrow and the gamma-ray observations made by Fermi and VERITAS are not 
simultaneous. We analysed therefore two alternative solutions focusing first
on the {\it Swift} and Fermi observations and then using the {\it Swift} 
and VERITAS data with different assumption about the peak positions.

To estimate the basic physical parameters of RGB J0710+591 we assume
both the position of the synchrotron peak ($\nu_s=7 \times 10^{17}$ Hz) and
the emission level at the peak ($\nu_s F_s(\nu_s) = 3 \times 10^{-11}$ 
erg cm$^2$ s$^{-1}$). It is also necessary to assume the frequency 
of the IC peak ($\nu_c=9.5 \times 10^{25}$ Hz) and the emission level 
at this peak ($\nu_c F_c(\nu_c) = 10^{-11}$ erg cm$^2$ s$^{-1}$).
However, for this peak we assume additional discrepancy
of about $\pm 30 \%$ in both the position and the emission level as well.
This discrepancy is introduced because of the unknown extragalactic 
absorption. The spectral index of the synchrotron emission below the 
peak is assumed to be $\alpha_1=0.5$ ($n_1=2$), whereas above the 
peak we use $\alpha_2=3/2$ ($n_2=4$).

\begin{figure}
\centering
\includegraphics[width=9cm]{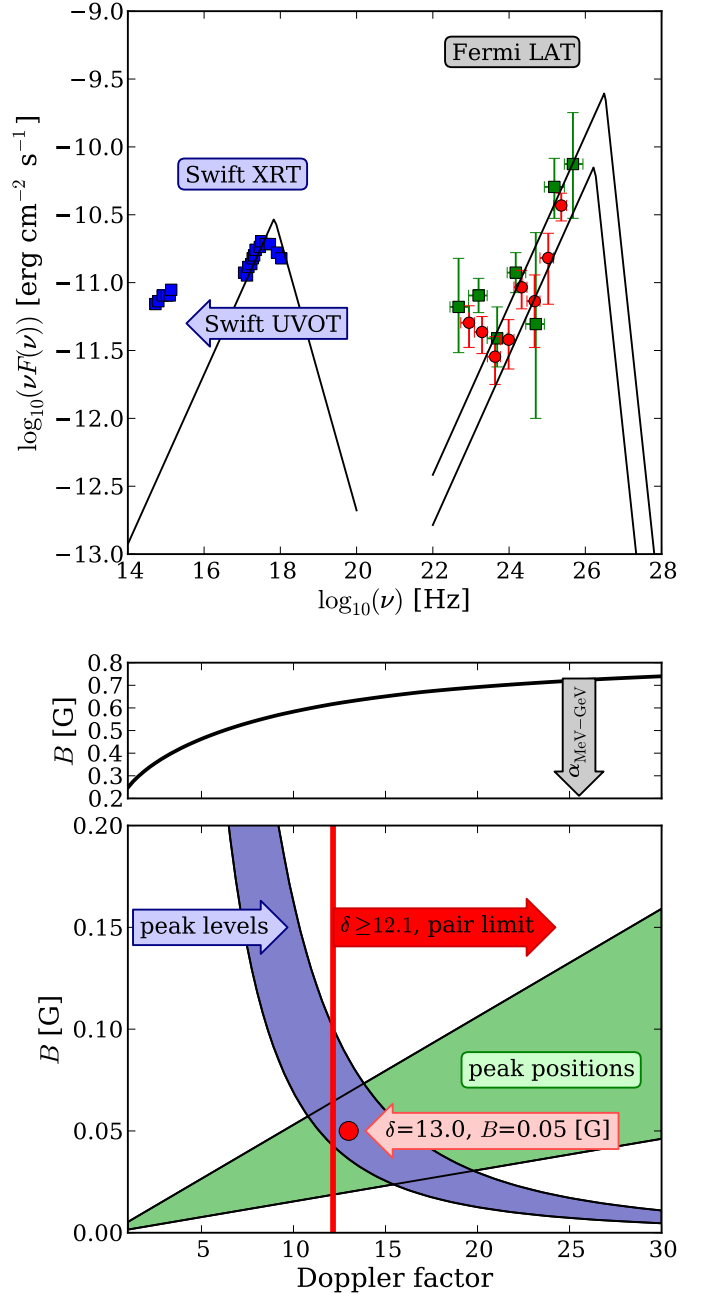}
\caption{The upper panel the shows multi--frequency emission of 1ES 0502+675
         observed by the {\it Swift} \& Fermi instruments 
         (Abdo et al. \cite{Abdo10b} -- squares, Abdo et al. 
         \cite{Abdo10c} -- circles) and the 
         approximate spectra used for
         the estimation presented in the lower panel.
         The values of $B$ \& $\delta$ selected for our 
         modelling are indicated by the dot.
        }
\label{fig_1es_est}
\end{figure}

With the above assumptions, we can estimate the values of $B$ and 
$\delta$ in the four different ways described in the previous 
subsections. The spectra used for the estimation and the
estimation result are presented in Fig. \ref{fig_rgb_est}. 
On the basis of the constraints obtained from the peak positions 
(Eq. \ref{eq_est_pek}) and the emission levels (Eq. 
\ref{eq_est_lev}), we found two crossing areas because
of the discrepancy in the IC peak. The two other
methods provide upper limit to the magnetic field
value and the lower limit to the Doppler factor.

To calculate some of the limiting curves, we have to determine 
the radius of the emitting region. This parameter in principle
can be constrained from the observed variability time scales
$R \lesssim c t_{\rm var} \delta / (1+z)$. However, there
is no information about the variability of RGB J0710+591.
We therefore selected as large as possible a value of
the radius $R=2 \times 10^{16}$ cm, which still provides
a good agreement for the four different estimation 
methods.

In the second approach, we focus on the VERITAS observations 
of RGB J0710+591 assuming that the IC peak is at the frequency 
$\nu_c=2 \times 10^{26}$ Hz and the emission level at this peak 
is $\nu_c F_c(\nu_c) = 9.5 \times 10^{-12}$ erg cm$^2$ s$^{-1}$. 
In addition me must assume that the synchrotron peak is 
placed above the observed range ($\nu_s=3 \times 10^{18}$ Hz 
and $\nu_s F_s(\nu_s) = 4 \times 10^{-11}$ erg cm$^2$ s$^{-1}$). 
This assumption is similar to the solution proposed by Acciari 
et al. (\cite{Acciari10}) that helps us to explain a ``flat'' 
($\Gamma \sim 2$) intrinsic TeV spectrum.

In the case of 1ES 0502+675, we assumed the same parameters for
the synchrotron peak as for RGB J0710+591 and that $\nu_c= 2.5 \times 10^{26}$ Hz, 
and $\nu_c F_c(\nu_c) = 1.5 \times 10^{-10}$ erg cm$^2$ s$^{-1}$ for the IC peak.
Moreover, we used $\alpha_1= 0.375$ ($n_1=1.75$) and $\alpha_2=2$ ($n_2=5$).
We explain this particular choice in the next section.

The main difference between this estimation and the calculations performed 
for RGB J0710+591 is in the upper limit to the magnetic field obtained
from the spectral slope in the MeV--GeV range. This estimate was 
calculated under the assumption that $x_B\simeq 1$ ($U_{\rm syn} \simeq U_B$). 
However, the observations indicate that the emission at the IC peak, in
this particular source, might even be one order of magnitude higher 
than the emission at the synchrotron peak. This would indicate that
$x_B >> 1$ but a more precise value of this parameter cannot be estimated 
from the observations because the IC scattering at the peak is
already in the Klein-Nishina regime. This illustrates a limitation 
of this particular estimation method. Finally, we selected as large as 
possible radius ($R=10^{16}$ cm), which provided a good agreement 
between our four different methods of the estimation.

\section{Modelling}

The information that we can obtain from the analysed observations is
significantly limited. We therefore decided to apply quite simple
model of the emission that is  directly related to our estimation methods.
The basic assumptions of the model that we adopted are identical to the 
assumptions made for the estimations. The VHE originates in a spherical source
uniformly filled by relativistic particles and an entangled magnetic 
field. The source evolves in time, ``fresh'' particles are 
continuously injected into the source where they lose
energy by means of the synchrotron and the IC emission. This is a
simple adaptation of the jet internal shock model, where the 
particles efficiently accelerated at a shock front fill
the downstream region of the shock. 

The evolution of the particle energy spectrum inside the source is
described by the kinetic equation
\begin{equation}
\frac{\partial N(\gamma,t)}{\partial t} + \frac{\partial }{\partial \gamma} 
\left[ \left\{ \dot{\gamma}_{\rm syn} + \dot{\gamma}_{\rm IC}\right\} N(\gamma,t) \right] = Q(\gamma) + P(\gamma, t),
\end{equation}
where $ \dot{\gamma}_{\rm syn} = - 4 \sigma_T c \gamma^2 U_B / (3 m_e c^2)$ 
is the synchrotron cooling rate and the IC cooling rate ($\dot{\gamma}_{\rm IC}$) 
is calculated according to the prescription of Sauge \& Henri (\cite{Sauge04})
for an isotropic distribution of soft photons, using the full Klein-Nishina 
cross--section in the head-on approximation (Jones \cite{Jones68}). The 
injection rate is described by a power-law function
\begin{equation}
Q(\gamma)= Q_0 \gamma^{-n_i}  \;\;\; {\rm for} \;\;\; \gamma \geq \gamma_i,
\end{equation}
where $\gamma_i \gg 1$ describes the minimum energy of the injected
particles. Finally, we simulate the pair creation process within the 
source, where the pair injection rate $P(\gamma,t)$ is calculated 
using the approach described by Sauge \& Henri \cite{Sauge04}. 
This process in principle should be negligible when we use the parameters
obtained from our estimations. However, in the estimation our intention 
was to use as large as possible a radius that gives a relatively small value
of the Doppler factor. The value is only slightly larger than the
lower limit obtained from our simple estimation (Eq. \ref{eq_par_lim}).
Therefore, the pair creation process may have a small impact 
on the evolution of the particle energy spectrum in our more precise
simulation. To solve the kinetic equation, we used the numerical
method described by Chiaberge \& Ghisellini \cite{Chiaberge99}.
Calculating spectra for each time step, we are carefully checked 
the energetic balance between the injected and radiated energy.

\begin{table}
\caption{The parameters used for the modelling}
\label{tab_mod_par}
\begin{tabular}{llll}
\hline
 & RGB J0710+591 & 1ES 0502+675 & \\
\hline
          & 1st (2nd) estimation       &                      & \\
$\delta$  & $8 \; (10)$                & $13$                 & \\
$B$       & $0.15 \; (0.13)$           & $0.05$               & G\\
$R$       & $2 \; (1) \times 10^{16}$  & $10^{16}$            & cm \\
$Q_0$     & $0.25 \; (3.5) \; \times 10^{5}$    & $4.5 \times 10^{11}$ & cm$^{-3}$ s$^{-1}$ \\
$\gamma_i$& $0.4 \; (1) \; \times 10^{6}$ & $4.5 \times 10^{5}$  & \\
$n_i$     & $3 \; (3)$                 & $4$ & \\
\hline
\end{tabular}
\end{table}

Such simple time--dependent modelling has been already proposed many 
times to explain evolution of the TeV blazars emission (e.g. Dermer 
et al. \cite{Dermer99}, Kusunose et al. \cite{Kusunose00}, B\"ottcher 
\& Chiang \cite{Bottcher02}, Sauge \& Henri \cite{Sauge04}). The main 
difference between our approach and the other models 
is in the minimum injected energy, where
in the other approaches $\gamma_i = 1$ is usually assumed.
In other words, we assume that the acceleration process at 
the shock front is very efficient and ``pushes'' almost all
particles towards high relativistic energies. This is somehow
similar to the formation of either a thermal or quasi-thermal 
particle energy distribution (e.g. Schlickeiser \cite{Schlickeiser84},
Katarzy\'nski et al. \cite{Katarzynski06b}).

To obtain an exceptionally hard spectrum in the MeV-GeV range, we needed 
to apply a cut-off to the low energy part of the spectrum. This cut-off can 
be produced in a natural way by the radiative cooling. Injecting
only high energy particles, we may obtain a broken power-law
particle distribution with the break at the $\gamma_i$ energy.
The minimum energy ($\gamma_m$) that the particles can reach 
in the source evolution time is given by Eq. \ref{eq_gam_min}.
(e.g.  Kardashev \cite{ Kardashev62}). This shows that
in principle, the particles might cool down to $\gamma \sim 1$,
if the source evolution is long enough. However, the minimum 
energy is inversely proportional to time, which means that the 
high energy particles ($\gamma \sim 10^5$) lose energy 
more rapidly than the medium energy particles ($\gamma \sim 10^3$). 
Moreover, in the downstream region of the shock there may be
another acceleration process, for example a turbulent second-order
Fermi acceleration. This relatively weak process may not produce
any high energy activity but it can be strong enough to compensate 
at some point for the radiative cooling and to keep the minimum energy 
significantly above the particle rest energy. This is, however,
a much more complex approach that requires an analysis of possible
activity. With no information about time--dependent flux 
variations, we simply assume that the source evolution time 
is just $2 R/c$.

The spectral index of the particle energy spectrum below the break will 
be constant $n_1=2$, whereas above the peak the index depends on 
the injection slope $n_2=n_i+1$. This, however, is true only for the 
radiative cooling caused by either the synchrotron emission or the IC 
emission in the Thompson regime.
In the Klein-Nishina regime, the scattering efficiency of
the high energy particles is significantly lower. This 
modifies the high energy part of the energy distribution 
(e.g. Moderski et al. \cite{Moderski05}). It is difficult 
to describe this effect analytically in our particular
scenario. We may perform only a simple estimation 
assuming that the energy density of the synchrotron 
radiation field is given by
\begin{equation}
U_{\rm syn} \simeq \frac{4 \pi}{c} \int_{\nu'_1}^{\nu'_2} I_s(\nu') d \nu',
\end{equation}
where $\nu'_1$ is the minimum frequency of the scattered photons
and $\nu'_2 = \min [ \nu'_x, 3 m_e c^2/(4 h \gamma)]$ introduces 
the KN restriction in a simple form (e.g. Chiaberge \& Ghisellini 
\cite{Chiaberge99}). Neglecting the lower boundary in the
above integral and assuming that $\nu'_2 = 3 m_e c^2/(4 h \gamma)$ 
and $I_{s} \propto (\nu')^{-\alpha_1}$, we may approximate
the IC cooling ratio as $\dot{\gamma}_{\rm IC} 
\propto - \gamma^{2-\alpha_1}$. Hence, the stationary
($\dot{N} = 0$) solution of the kinetic equation is given by
$N \propto \gamma^{-{2-\alpha_1}}$, where the slope 
is significantly lower than the value $N \propto \gamma^{-2}$
obtained for a constant classical synchrotron or
IC cooling. This agrees with our more precise numerical 
tests, which indicate that the radiative cooling dominated 
by the IC scattering in the KN regime can reduce the
particle energy slope by a factor $\lesssim 1/2$. This appears 
to be true for the cooled (below the break) and 
injected (above the break) part of the energy spectrum. 
Since the emission of 1ES 0502+675 seems to be 
significantly affected by the scattering in the
KN regime, we assume that $n_1=1.75$ for the estimation 
made in the previous section for this source. We note
that the energy spectrum index less than two helps
in addition to explain exceptionally hard MeV-GeV 
spectra.

\begin{figure}
\centering
\includegraphics[width=9cm]{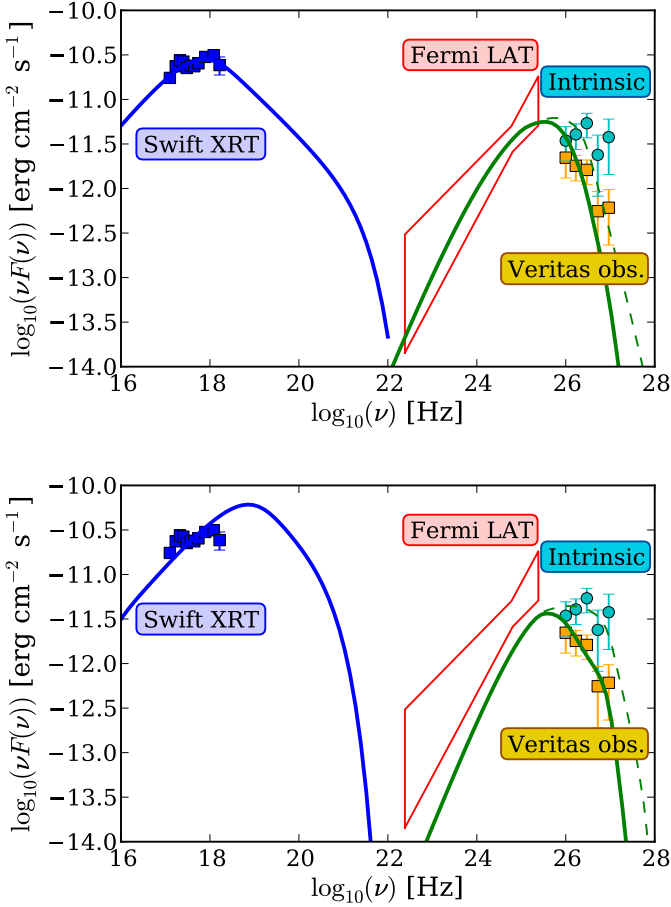}
\caption{The modelling of the RGB J0710+591 spectral emission. The upper panel
         shows the first approach where we considered the {\it Swift} and Fermi
         observations only. In the lower panel, we demonstrate an alternative approach
         that can explain the origin of the TeV emission observed by VERITAS. Note that
         the Fermi and the VERITAS observations were not simultaneous.
         To obtain the observed IC spectra, we used the EBL lower-limit model 
         proposed by Kneiske \& Dole (\cite{Kneiske10}), where the dashed lines 
         show the intrinsic unabsorbed emission.
        }
\label{fig_RGB_mod}
\end{figure}

Figure \ref{fig_RGB_mod} shows the modelling of the RGB J0710+591 emission.
We present the results obtained for two different sets of physical parameters
derived from the estimations performed in the previous section. Note that the IC
spectrum obtained in the first case is unable to explain the intrinsic shape of 
the emission. In the second case, the intrinsic spectrum is reproduced well but the 
level of the MeV-GeV emission is significantly lower than the observed level. However,
the Fermi and the VERITAS observations are not simultaneous, hence there may
be a difference in the emission levels.
The detailed values of the physical parameters used for the modelling 
are given in Tab.~\ref{tab_mod_par}. There are only six important parameters  
illustrating the simplicity of the model. In addition, we assumed that the 
maximum Lorentz factor of the injected particles is $\gamma>10^7$ and 
the source evolution time $t_{\rm evo} = 2 R/c$.

\begin{figure}
\centering
\includegraphics[width=9cm]{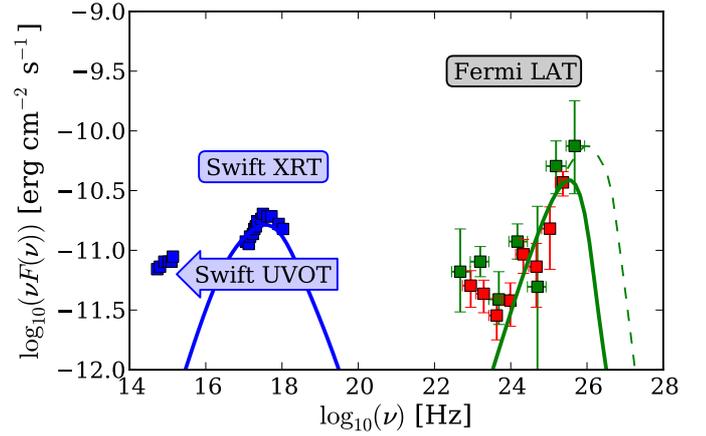}
\caption{The spectra obtained from the modelling of 1ES 0502+675 for
         the physical parameters obtained from our estimation.
        }
\label{fig_1ES_mod}
\end{figure}

For 1ES 0502+675, we do not attempt to reproduce the {\it Swift} UVOT observations.
This part of the emission might be dominated by either the host galaxy or more 
extended jet structures as demonstrated for example in the case of Mrk 501 
(Katarzynski et al. \cite{Katarzynski01}). Moreover, no direct correlation between 
X-ray and optical variability in other TeV blazars also implies that 
the optical-UV emission has a different origin. Our modelling is also unable to explain 
the inverted spectral slope in the 100 MeV to 1 GeV energy range. This is an unexpected
and puzzling feature of the emission that will require additional investigation. 
However, this may be simply the IC emission produced by the extended jet structures,
which also dominate the emission in the optical-UV range.

\section{Summary}

Several blazars observed by the Fermi Gamma-Ray Space Telescope have exceptionally
hard spectra in the MeV-GeV range. We have demonstrated that in principle all sources 
with a spectral index $\alpha \lesssim 0.65$ should be classified as 
sources with an exceptionally hard MeV-GeV spectrum. Therefore, we have attempted
to reproduce these spectral slopes by assuming a low energy cut-off in the 
particle energy distribution. This approach had been previously developed to 
explain hard spectra in the TeV range and at least in one case  
(1ES 0229+200 Tavecchio et al. \cite{Tavecchio09}) it was found to be correct. 

We derived four different methods to estimate the
basic parameters of the emitting region. Three of these
methods had been previously proposed and we simply adapted them 
to the case of the low energy cut-off. Our method that gives
an upper limit to the magnetic field, derived from the spectral
slope in the MeV-GeV range, was proposed here for the first time. 
This particular constraint can be very useful but it requires
accurate measurements of the spectral index in the not-absorbed 
MeV-GeV range.

Finally, we used our modelling estimates that succeed to
explain the exceptionally hard spectra in terms of
particle energy evolution. Our model is relatively simple but 
more complex, and hence more accurate modelling
would require far tighter observational constraints. We have attempted 
to explain the emission of two different TeV blazars. The particle
evolution in RGB J0710+591 seems to be dominated by
synchrotron cooling, whereas in 1ES 0502+675 IC cooling appears to
dominate. In both cases, the estimated
parameters of $B$ \& $\delta$ provide good results in 
our modelling. This type of modelling is crucial for understanding
the nature of the blazar emission process, especially in 
non--standard cases such as the exceptionally hard MeV-GeV 
emission. One interpretation of exceptionally hard spectra is 
that for some reason only high energy particles can produce the 
X-ray and gamma-ray activity of TeV blazars.

\begin{appendix}

\section{Approximation of synchrotron emission}

For a power-law particle energy distribution 
\begin{equation}
N(\gamma) = K \gamma^{-n}, \;\; [{\rm cm}^{-3}]
\end{equation}
the synchrotron emissivity can be approximated by 
a simple power-law function
\begin{equation}
j_s(\nu') = \frac{1}{4 \pi} C(n) K B^{\alpha+1} (\nu')^{-\alpha}, \; [{\rm erg} \; {\rm s}^{-1} \; {\rm cm}^{-3} {\rm sterad}^{-1} {\rm Hz}^{-1}],
\label{eq_syn_emi}
\end{equation}
where $\alpha = (n-1)/2$ and
\begin{eqnarray}
C(n) & = & \frac{4 \sqrt{3} \pi e^3}{m_e c^2} \left( \frac{3e}{2\pi m_e c} \right)^{\frac{n-1}{2}} \Gamma\left(\frac{3n-1}{12}\right) \Gamma\left(\frac{3n+19}{12}\right)\nonumber\\
     & \times & \Gamma \left( \frac{n+5}{4}\right)  /  \left( 8 \sqrt{\pi} (n+1) \Gamma((n+7)/4) \right)
\end{eqnarray}
(e.g. Ginzburg \& Syrovatskii \cite{Ginzburg65}). Assuming an optically
thin source, we can write the synchrotron luminosity $L_s=4 \pi V j_s$ [erg],
where $V$ is the source volume. If our source is spherical then the intensity 
of the surface emission is given by 
\begin{equation}
I_s = \frac{4}{3} j_s R \; [{\rm erg} \; {\rm s}^{-1} \; {\rm cm}^{-2} {\rm sterad}^{-1} {\rm Hz}^{-1}],
\label{eq_syn_int} 
\end{equation}
where $R$ is the source radius.
This gives $L_s = 4 \pi^2 R^2 I_s$ and using the standard luminosity--to--flux
conversion, we can obtain the observed flux $F_s = \pi (R/D_L)^2 I_s$ 
[erg cm$^{-2}$], where $D_L$ is the luminosity distance. For relativistically 
moving sources at cosmological distances, we have to apply additional
transformations to the flux 
\begin{equation}
F_s(\nu) = \pi \frac{R^2}{D_L^2} (1+z) \delta^3 I_s(\nu'),
\label{eq_flx_tra}
\end{equation} 
as well as the frequency $\nu = \nu' \delta / (1+z)$. 

\begin{figure*}[!t]
\centering
\includegraphics[width=18cm]{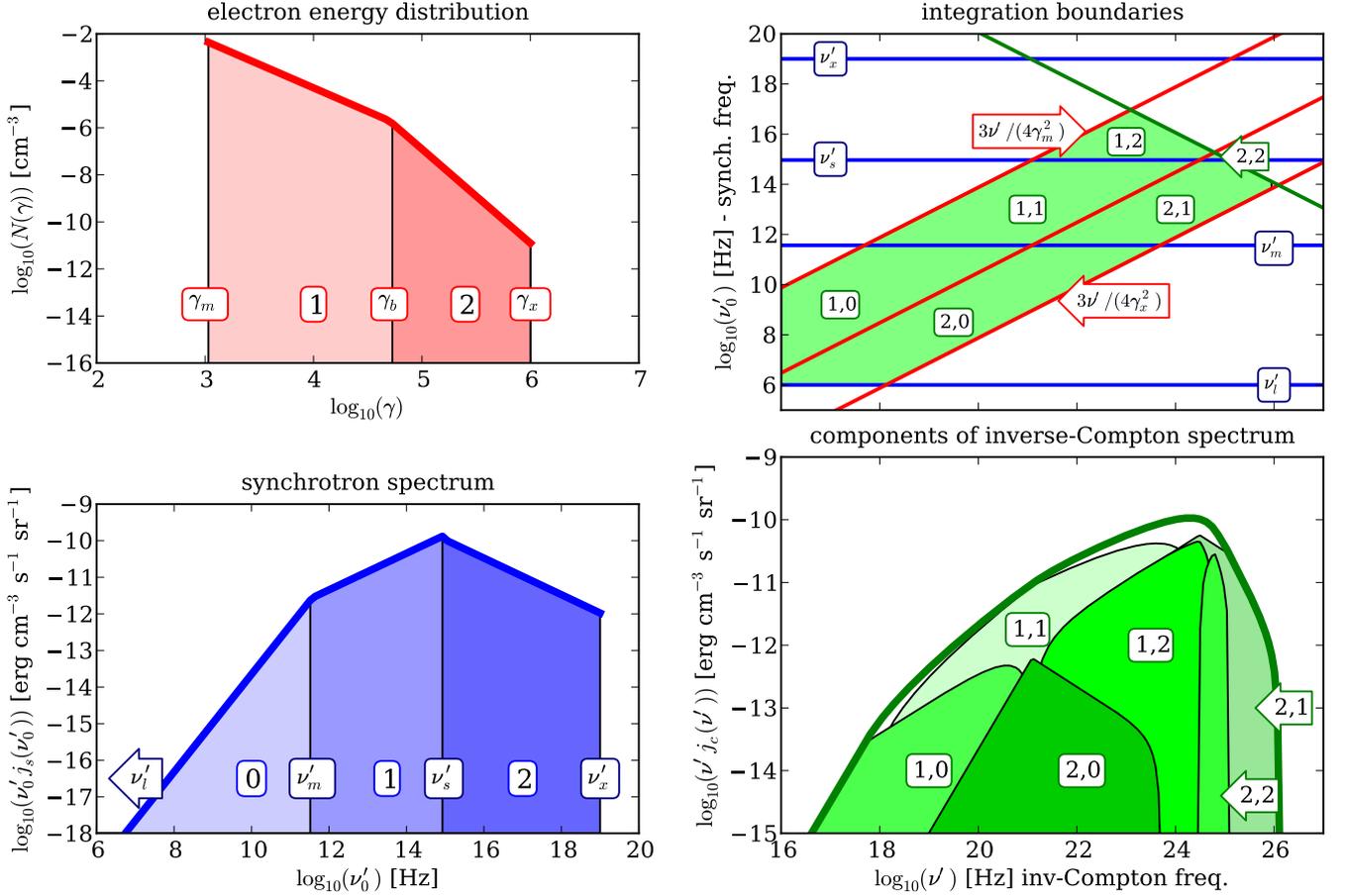}   
\caption{An example of an approximation of SSC emission. The figure shows a broken power-law ($\gamma_b$) 
electron spectrum with a low energy cut-off ($\gamma_m$) that extends up to some maximal energy ($\gamma_x$). This
distribution generates synchrotron emission with two breaks ($\nu'_m, \nu'_s$) and extends from some minimal
($\nu'_l$) to some maximal frequency ($\nu'_x$). The energy spectrum is divided into two parts, whereas
the synchrotron radiation is divided into three parts. This provides six different components of the inverse-Compton emission.
Each component must be integrated with different boundary conditions. The upper right panel demonstrates how to
choose these conditions. This panel shows synchrotron frequencies as functions of 
the IC photon frequencies, where the horizontal lines indicate the characteristic frequencies in the synchrotron 
spectrum. Three slanted lines are calculated according to the well--know relation $\nu'_{\rm IC} = (4/3) \gamma^2 
\nu'_{\rm syn}$ for the three characteristic particle energies ($\gamma_m, \gamma_b, \gamma_x$). The single 
opposite slanted line shows the Klein-Nishina restriction according to the approximation $x'= 3 / (4 x'_0)$. The
crossing lines are produce six rectangle areas that indicate the lower and upper integration boundaries for each
component. Each component of the IC emission is indicated by a pair of numbers where the first number 
refers to the part of the electron energy spectrum and the second number shows the part of the radiation field 
used to create this component.
}
\label{fig_com_emi}
\end{figure*}

\section{Approximation of inverse-Compton emission}

The inverse-Compton emissivity can be approximated by a simple 
formula
\begin{equation}
j_c(x') = \frac{\sigma_T}{4} \int^{x_2}_{x_1} N \left( \sqrt{\frac{3 x'}{4x'_0}}\right)\sqrt{\frac{3 x'}{4x'_0}} \frac{I_s(x'_0)}{x'_0} d x'_0,
\label{eq_com_emi}
\end{equation}
where $x'=h\nu'/{m_e c^2}$ is the inverse-Compton photon energy 
normalized to the electron rest energy and $x'_0$ is the energy 
of a synchrotron photon also divided by $m_e c^2$. The particle 
energy spectrum, in our particular estimations, is assumed to be 
a broken power-law 
\begin{equation}
N(\gamma) = \left\{
             \begin{tabular}{ll}
              $K_1 \gamma^{-n_1}$     & ${\rm for}~~ \gamma_m \le \gamma \le \gamma_b$ \\
              $K_2 \gamma^{-n_2}$     & ${\rm for}~~ \gamma_b   < \gamma \le \gamma_x$
             \end{tabular}
            \right. ,
\label{eq_bpl_esp}
\end{equation}
where $K_2= K_1 \gamma^{n_2-n_1}$ and $\gamma_{\rm m} >> 1$. For such an energy
spectrum, we can approximate the synchrotron intensity by a double broken 
power-law function
\begin{equation}
I_s(x'_0) = \left\{
             \begin{tabular}{ll}
              $ s_0 (x'_0)^{-\alpha_0}$ & for $x'_l \le x'_0 \le x'_m$\\
              $ s_1 (x'_0)^{-\alpha_1}$ & for $x'_m   < x'_0 \le x'_s$\\
              $ s_2 (x'_0)^{-\alpha_2}$ & for $x'_s   < x'_0 \le x'_x$\\
             \end{tabular}
            \right. ,
\label{eq_rad_fie}
\end{equation}
that extends from some minimal energy ($x'_l$) up to the maximal energy ($x'_x$) and contains
breaks at the characteristic energies ($x'_m, x'_s$) related to the particle
energy by
\begin{equation}
x'_{m/s} = \frac{h}{m_e c^2} 3.7 \times 10^6 \gamma^2_{m/b} B.
\end{equation}
According to the Eqs \ref{eq_syn_emi} and \ref{eq_syn_int}, 
the normalizing coefficients in the synchrotron intensity are given by
\begin{eqnarray}
 s_{1/2} & = & \frac{1}{4 \pi} C(n_{1/2}) \left(\frac{m_e c^2}{h}\right)^{-\alpha_{1/2}} R K_{1/2} B^{\alpha_{1/2} + 1},\\
 s_0     & = & s_1 (x'_m)^{\alpha_0 -\alpha_1}.
\end{eqnarray}
Note that inside the source the intensity of the synchrotron radiation field is on 
average $I_s \simeq (3/4) R j_s$ (e.g. Kataoka et al. \cite{Kataoka99}). Therefore, 
there is no constant $4/3$ in the above formulae.

Since the energy spectrum (Eq. \ref{eq_bpl_esp}) and the radiation field spectrum 
(Eq. \ref{eq_rad_fie} ) are divided into the simple power-law functions, we can
split the inverse-Compton emissivity into six components where each component is
described by
\begin{equation}
j_{(a,b)}(x') = \frac{\sigma_T}{4} s_b K_a \left( \frac{3 x'}{4}\right)^{\frac{1-n_a}{2}} 
          \int^{x_2}_{x_1} (x'_0)^{\frac{n_a-1}{2}-1-\alpha_b} d x'_0,
\end{equation}
where index $a$ (equal either to 1 or 2) describes parts of  the particle energy spectrum and 
index $b$ (equal 0, 1 or 2) indicates parts of the synchrotron spectrum. Depending
on the $a, b$ values, this gives two different results. For $a=b$, we have
\begin{equation}
j_{(a,a)}(x') = \frac{\sigma_T}{4} s_a K_a \left( \frac{3 x'}{4}\right)^{-\alpha_a} \ln(x_2/x_1)
\label{eq_com_aa}
\end{equation}
and for $a \ne b$
\begin{equation}
j_{(a,b)}(x') = \frac{\sigma_T}{4} s_b K_a \left( \frac{3 x'}{4}\right)^{-\alpha_a}
                \frac{x_2^{\alpha_a} - x_1^{\alpha_b}}{\alpha_a -\alpha_b}.
\label{eq_com_ab}
\end{equation}
Finally, we have to define the integration boundaries ($x_1, x_2$), which are different 
for each component. In general, we can define the lower integration boundary as 
a maximum of two energies of synchrotron photons
\begin{equation}
x_1 = \max(x'_p, x'_q),
\end{equation}
where $x'_p$ is equal to $x'_l, x'_m$,  or  $x'_s$ for $b=0, 1, 2$, respectively, and
$x'_q = (4/3) \gamma^2 x'_0$ is calculated for the characteristic particle energies 
$\gamma_b, \gamma_x$ for $a=1$ and 2, respectively. The upper boundary condition is 
calculated to be the minimum of three energies
\begin{equation}
x_1 = \min \left(x'_p, x'_q, \frac{3}{4x'} \right),
\end{equation}
where $x'_p$ is equal to either $x'_m, x'_s$,  or  $x'_x$ for $b=0, 1, 2$, respectively
and $x'_q$ is calculated for $\gamma_m, \gamma_b$ for $a=1$ and 2, respectively.
The last part of this maximum comes from the simple approximation of the 
Klein-Nishina limit where
\begin{equation}
x' = \frac{4}{3} \gamma^2 x'_0.
\end{equation}
An example spectrum of the approximate IC emission and a description of the
integration boundaries and the components of the spectrum is presented in Fig. 
\ref{fig_com_emi}. This particular spectrum was calculated for a set of the 
parameters that gives all six components of the spectrum. Note that for the 
parameters that are typical for TeV blazars, the IC scattering occurs predominantly 
in the KN regime and only two components ($j_{(1,1)}$) and ($j_{(2,1)}$)
creates the spectrum if $1 << \gamma_m << \gamma_b$. In the case where 
$1<< \gamma_m \simeq \gamma_b$, the first part of the energy spectrum ($a=1$) 
and the second part of the radiation field ($b=1$) basically do not exist.
In such a case, the spectrum is dominated by the $j_{(2,0)}$ component with the
spectral index in the MeV-GeV range $\alpha_0 = -1/3$. However, this is an
extreme limiting case, never directly observed.

\end{appendix}

\begin{acknowledgements}

We thank the anonymous referee for a careful review and a helpful report.

\end{acknowledgements}


\begin{thebibliography}{}
\bibitem[2010a]{Abdo10a}   	Abdo, A. A., Ackermann, M., Ajello, M., et al.             2010a, ApJS, 188, 405
\bibitem[2010b]{Abdo10b}        Abdo, A. A., Ackermann, M., Agudo, I., et al.              2010b, Apj, 716, 30
\bibitem[2010c]{Abdo10c}   	Abdo, A. A., Ackermann, M., Ajello, M., et al.             2010c, ApJ, 710, 1271
\bibitem[2010]{Acciari10}       Acciari, V. A., Aliu, E., Arlen, T., et al.                2010., ApJ, 715, L49    
\bibitem[2006]{Aharonian06}   Aharonian, F., Akhperjanian, A. G., Bazer-Bachi, A. R., et al.  2006, Nature, 440, 1018
\bibitem[2007a]{Aharonian07a} Aharonian, F., Akhperjanian, A. G., Bazer-Bachi, A. R., et al. 2007a, ApJ, 664, 71
\bibitem[2007b]{Aharonian07b} Aharonian F., Akhperjanian, A. G.; Barres de Almeida, U. et al.,            2007b, A\&A, 475, L9
\bibitem[2009]{Atwood09}      Atwood, W. B., Abdo, A. A., Ackermann, M., et al.             2009, ApJ, 697, 1071
\bibitem[2002]{Bottcher02}    B\"ottcher, M., \& Chiang, J.                                2002, ApJ, 581, 127
\bibitem[1996]{Bloom96}       Bloom, S. D. \& Marscher, A. P.                              1996, ApJ, 461, 657
\bibitem[1997]{Catanese97}    Catanese, M., Bradbury, S., M., Breslin, A., C., et al.      1997, ApJ   , 487, L143
\bibitem[1999]{Chiaberge99}   Chiaberge, M. \& Ghisellini, G.,                             1999, MNRAS, 306, 551
\bibitem[1990]{Coppi90}       Coppi, P. S. \& Blandford R. D.,                             1990, MNRAS,245, 453
\bibitem[2002]{DeJager02}      De Jager, O. C., \& Stecker, F. W.                           2002, 566, 738
\bibitem[1997]{Dermer97}      Dermer, C. D., Sturner, S. J. \& Schlickeiser, R.,           1997, ApJS, 109, 103
\bibitem[1999]{Dermer99}      Dermer, C. D., Li, H.,  \& Chiang, J.,                       1999, ApL\&C, 39, 1
\bibitem[1995]{Dondi95}       Dondi, L, \& Ghisellini, G.                                  1995, MNRAS, 273, 583
\bibitem[2008]{Fossati08}     Fossati, G., Buckley, J. H., Bond, I. H., et al.             2008, ApJ, 677, 906
\bibitem[2008]{Franceschini08} Franceschini A., Rodighiero G., \& Vaccari M.,              2008, A\&A, 487, 837
\bibitem[1998]{Ghisellini98}  Ghisellini, G., Celotti, A. Fossati, G., et al.              1998, MNRAS, 301, 451 
\bibitem[2010]{Ghisellini10}  Ghisellini, G., Tavecchio, F., Foschini, L., et al.          2010, MNRAS, 402, 497
\bibitem[1965]{Ginzburg65}     Ginzburg, V. L. \& Syrovatskii, S. I.                       1965, ARA\&A, 3, 297  
\bibitem[1996]{Inoue96}       Inoue, S. \& Takahara, F.,                                   1996, ApJ, 463, 555
\bibitem[1968]{Jones68}       Jones, F., C.,                                               1968, Phys. Rev., 167, 1159
\bibitem[2004]{Massaro04}      Massaro, E., Perri, M., Giommi, P., et al.                  2004, A\&A, 413, 489
\bibitem[1962]{Kardashev62}   Kardashev, N.S.,                                             1962, Soviet Astronomy--AJ, 6, 317
\bibitem[1999]{Kataoka99}      Kataoka, J., Mattox, J. R., Quinn, J., et al.               1999, ApJ, 514, 138
\bibitem[2001]{Katarzynski01} Katarzy\'nski, K., Sol, H., \& Kus, A.,                     2001, A$\&$A, 367, 809 
\bibitem[2006a]{Katarzynski06a} Katarzynski, K., Ghisellini, G., Tavecchio, F., et al.      2006a, MNRAS, 368, 52
\bibitem[2006b]{Katarzynski06b} Katarzynski, K., Ghisellini, G., Mastichiadis, A., et al.   2006b, A\&A, 453, 47
\bibitem[2004]{Kneiske04}     Kneiske, T. M., Bretz, T., Mannheim, K., \& Hartmann, D. H., 2004, A\&A, 413, 807
\bibitem[2010]{Kneiske10}     Kneiske, T. M., \& Dole, H.                                 2010, A\&A, 515, A19
\bibitem[2000]{Kusunose00}     Kusunose, M., Takahara, F., \& Li, H.,                     ApJ, 536, 299
\bibitem[1997]{Mastichiadis97}Mastichiadis, A., \& Kirk, J.,                              1997, A\&A, 320, 19
\bibitem[2005]{Moderski05}     Moderski, R., Sikora, M., Coppi, P., S., Aharonian, F.,     2005, MNRAS, 363, 954
\bibitem[2006]{Nieppola06}     Nieppola, E.; Tornikoski, M., \& Valtaoja, E.               2006, A\&A, 445, 441
\bibitem[2009]{Ong09}          Ong, R. A.                                                  2009 ATel, 2301, 1O
\bibitem[2004]{Sauge04}       Sauge, L., \& Henri, G.,                                     2004, ApJ, 616, 136
\bibitem[1984]{Schlickeiser84} Schlickeiser, R.                                           1984, A\&A, 136, 227
\bibitem[1998]{Tavecchio98}   Tavecchio, F., Maraschi, L. \& Ghisellini, G.,              1998, ApJ, 509, 608 
\bibitem[2009]{Tavecchio09}   Tavecchio, F., Ghisellini, G., Ghirlanda, G., et al.        2009, MNRAS, 399, 59

\end{thebibliography}
\end{document}